\documentclass[]{ar}
\usepackage[english]{babel}
\usepackage{psfig,graphicx}
\usepackage{supertab}

\onecolumn
\begin{document}

\title{Optical spectrum of the  post-AGB star HD\,56126 \newline in the
        region 4010--8790\,\AA}

\author{V.G.~Klochkova$^1$, E.L.~Chentsov$^1$, N.S.~Tavolganskaya$^1$,
         and M.V.~Shapovalov$^2$}

\date{\today}	     

\institute{Special Astrophysical Observatory RAS, Nizhnij Arkhyz,  369167 Russia, 
           \and Rostov State University, Rostov-on-Don, Russia}

\abstract{We studied in detail the optical spectrum of the post-AGB
star HD\,56126 associated with the IR-source IRAS\,07134$+$1005. We use
high resolution spectra (R\,=\,25000 and 60000) obtained with the echelle
spectrographs of the 6-m telescope. About one and a half thousand
absorptions of neutral atoms and ions, absorption bands of C$_2$, CN, and
CH molecules, and interstellar bands (DIBs) are identified in the 4010 to
8790\,\AA\AA{} wavelength region, and the depths and radial velocities of
these spectral features are measured. Differences are revealed between the
variations of the radial velocities measured from spectral features of
different excitation. In addition to the well-known variability of the
H$\alpha$ profile, we found variations in the profiles of a number of
Fe{\sc ii}, Y{\sc ii}, and Ba{\sc ii} lines. We also produce an
atlas\footnote{The full version of the Atlas is available in electronic
form from: {\it http://www.sao.ru/hq/ssl/Atlas/Atlas.html}} of the
spectrum of HD\,56126 and its comparison star $\alpha$\,Per.}

\authorrunning{Klochkova et al.}
\titlerunning{Spectroscopy of HD56126}

\maketitle

\section{Introduction}

The star HD\,56126 is observed in the post-AGB phase of its evolution.
While undergoing this short-lived stage (according to Bl\"ocker
[\cite{Block}], this phase lasts for $\Delta T \approx 10^3\div10^4$\,years),
the star passes to the stage of a planetary nebula and therefore post-AGB
stars are also commonly referred to as ``protoplanetary nebulae'' (PPNe).
In the Hertzsprung--Russell diagram post-AGB stars move at almost constant
luminosity leftward of the AGB and become increasingly hotter. These
objects, which are descendants of AGB-stars, can be used to trace the
physical and chemical parameters of the interstellar matter due to a
change in the energy source, which is accompanied by a change of the star
structure, ejection of the envelope, and mixing of matter.

The main task of our research is to reveal the chemical composition
anomalies that are due to the nuclear synthesis of chemical elements in
the interiors of low- and intermediate-mass stars (less than
8--9\,M$_{\odot}$) and subsequent dredge-up of the products of synthesis
to the surface layers of stellar atmospheres. We use our high-precision
spectroscopic data not only to study the chemical composition, but also to
perform a detailed analysis of the velocity field in the atmospheres of
these stars, which constitutes a separate astrophysical problem.
In addition, the high quality observational data allowed us to produce an
atlas of the spectrum of a typical post-AGB star over a wide wavelength
interval. To this task, we chose the supergiant star HD\,56126
($Sp$=$F5Iab$), which is the optical component of the IR source
IRAS07134\,+\,1005 with a double-peaked spectral energy distribution (SED)
typical of PPN. The star HD\,56126 is located outside the galactic plane,
its galactic coordinates are l=$206\lefteqn{.}^{\circ}75$,
b=$+9\lefteqn{.}^{\circ}99$. Note that HD\,56126 is a generally recognized
{\bf canonical object} in the phase of transition from the asymptotic
giant branch to a planetary nebula. In addition to the anomalous SED
mentioned above, which is due to the circumstellar dust, this star
exhibits other, highly conspicuous, features specific for this class of
objects [\cite{Kwok}]: the optical component of the PPN is an
$F5Iab$-type supergiant at a high galactic latitude; the central star is
surrounded by an extended nebula, which, according to HST observations
[\cite{Ueta}], has the largest angular size $\beta>4''$ among PPN objects of
this type; the optical spectrum exhibits variable complex
emission--absorption profile of the $H\alpha$ and shows spectral features
that are indicative of the current mass outflow. Based on their
high-resolution spectroscopy ($R$=860000, $FWHM$=0.35\,km/s) of HD\,56126,
Crawford and Barlow [\cite{Crawford}] revealed the multicomponent structure
of the K{\sc i} and C$_2$ features, which is indicative of repeated
episodes of mass ejection from the star.

Subsequent studies of HD\,56126 and of the associated IR source revealed a
number of properties, which allowed the object to acquire the canonical
status in its class. First, an analysis of the spectra obtained with the
echelle spectrograph of the 6-m telescope, allowed Klochkova [\cite{Kloch95}]
to conclude that HD\,56126 is a metal-poor star with
$[Fe/H]_{\odot}$=$-1.0$ and high excess of carbon and $s$-process
elements. Second, IRAS\,07134+1005 was found to belong to the group of
PPNe whose IR-spectra exhibit an emission feature at $\lambda$=21\,$\mu$m.
Objects of this small subgroup were found to exhibit a correlation between
the presence of this 21\,$\mu$m--feature and the manifestation of
products of stellar nucleosynthesis in the outer atmospheric layers:
overabundance of carbon and heavy metals of the $s$-process. This so far
unexplained correlation has been found independently by Klochkova
[\cite{Kloch98}] and a group of other authors [\cite{Decin}].
Thus HD\,56126 possesses the complete set of features peculiar to the entire
family of  PPNe, and this fact determines the importance of the detailed
spectroscopy of this object and preparation of an atlas of its optical spectrum
over a wide spectral region. This task is facilitated by HD\,56126 being
the brightest  ($B$=9$\lefteqn{.}^m11$, $V$=8$\lefteqn{.}^m27$) star among
carbon-rich PPNe and hence the most accessible star for high-resolution
spectroscopy among the objects of this type.

Section\,\ref{obs} gives a brief description of the methods of
observation and data reduction employed in this paper.
Section\,\ref{results} presents the peculiar features of the
spectrum of  HD\,56126, and section\,\ref{RV-var} describes the
field of radial velocities  $V_r$ in the atmosphere and envelope of the
star. We also briefly discuss the radial-velocity variability and the
variability of selected spectral-line profiles. Section\,\ref{atlas}
describes the spectroscopic atlas, identification of spectral features, and
compares the spectrum of HD\,56126 with that of the standard supergiant
$\alpha$\,Per ($Sp=F5Iab$).

\section{Observations and reduction of spectra}\label{obs}

We performed spectroscopic observations of HD\,56126 and $\alpha$\,Per
with the 6-m telescope of the Special Astrophysical Observatory. We obtained 
all spectra with NES [\cite{nes1,nes2}] and Lynx [\cite{lynx,lynx2}] echelle
spectrographs operating in the Nasmyth focus. A 2048$\times$2048 CCD and
image slicer [\cite{slicer}] with the NES spectrograph allows taking spectra with a
resolution of $R$\,$\approx$\,60000, whereas the Lynx spectrograph
equipped with a 1K$\times$1K CCD yields a resolution of
$R$\,$\approx$\,25000. The Table~\ref{RV} gives the dates of observations
and the spectral region recorded.

We use the modified ECHELLE context of MIDAS to  extract data from
two-dimensional echelle spectra  (see [\cite{ECHELLE}] for details).
Cosmic-hit removal was performed via median averaging of two successive
spectra. Wavelength calibration was made using Th-Ar-lamps. We use DECH20
[\cite{gala}] code to perform spectrophotometric and position measurements.
In particular, we determine the radial velocities from individual lines and
their components by superimposing the direct and mirror-reflected profiles. We
determine the position zero point for each spectrogram by referring it to the
positions of ionospheric emission features of the night sky and to those of
telluric absorptions, which show up against the spectrum of the object. The
accuracy of  {\bf  single line} velocity measurements in the spectra obtained
is better than 1.0 and 1.5\,km/s, for NES and Lynx spectrographs, respectively.

\begin{table*}[hbtp]
\caption{\footnotesize Log of observations  and results of $V_r$
        measurements. Column 4 gives the mean  $V_r$ averaged over weak lines
        ({\it r}\,$\rightarrow$\,1). For Fe{\sc ii}(42), H$\alpha$ and
        D lines of Na{\sc i} we give the velocities inferred from the
        positions of the strongest line components. The numbers in parentheses
        give the velocities inferred from weaker components. Slanted font in
        column~5 indicates the velocities inferred from the IR oxygen triplet
        O{\sc i}\,$\lambda$\,7773\,\AA{}. Semicolumn indicates uncertain data}
\medskip
\begin{tabular}{l|   l|  l|  l|  c|   l|  c|   l|  l|   l|  l|  l}
\hline
Date & Spectro- & $\Delta\lambda$, \AA{} & \multicolumn{7}{c}{\small $V_r$} \\  
\cline{4-12}
 & graph & &{\it r}\,$\rightarrow$\,1 & Fe{\sc ii}(42) & H$\beta$ &\quad H$\alpha$
                &\quad D\,Na{\sc i} & C$_2$ &\multicolumn{3}{c}{\small interstellar} \\ 
\hline
\quad  1&\quad 2&\qquad 3&4&5&6&7&\quad 8&9& 10& 11 & 12\\  
\hline
\multicolumn{12}{c}{\small \underline  {\hspace{0.5cm} HD\,56126 \hspace{0.5cm}}} \\ 
12.01.93&Lynx&5560--8790& 88.8 &{\it 91} &--  &78 (100:)&77      &--  &--   &-- &--  \\ 
10.03.93&Lynx&5560--8790& 89.0 &{\it 93} &--  &71 (43:) &75: &-- &--   &--   &--    \\ 
04.03.99&Lynx&5050--6640& 85.9 & 77      &--  &76 (43:) &78      &77.1&--   &--   &--    \\ 
20.11.02&NES &4560--5995& 89.6 &95 (80:) &89  &--       &75 (89) &77.2&12.0 &23.5 & 30.8\\ 
21.02.03&NES &5150--6660& 88.8 &96:      &--  &88 (112:)&75 (89) &77.1&12   &24 & 31     \\ 
12.04.03&NES &5270--6760& 88.4 &--       &--  &82 (103:)&75 (89:)&-- &13 &23   & 30.5 \\ 
14.11.03&NES &4518--6000& 85.3 &96 (87:) &97  &-- &75 (87:)&76.9&12.5 &--   &--     \\ 
10.01.04&NES &5270--6760& 86.7 & --      &--  &54:      &76 (86:)&--  &13.0 &23.5 & 31  \\ 
09.03.04&NES &5275--6767& 89.8 &--       &--  &58 (74:) &76 (89) &-- &13 &24   & 31   \\ 
12.11.05&NES &4010--5460& 82.5 &97 (77:) &98  &--       & --     &77.5&--    &--  &--  \\ 
\multicolumn{12}{c}{\small \underline{\hspace{0.8cm} $\alpha$\,Per \hspace{0.8cm}}} \\ 
04.03.99&Lynx&5050--6620&$-1.2$& $-1$    &--  &$-2$     &--      &--  &&--& \\ 
02.08.01&NES &3500--5000&$-1.8$& $-1:$   &$-2$&--       &--      &--  &&--& \\ 
11.11.05&NES &4010--5460&$-2.0$& $-2$    &$-2$&--       &--      &--  &&--&  \\ 
12.11.05&NES &4560--6010&$-1.9$& $-2$    &$-2$&--       &--      &--  &&--&  \\ 
\hline
\end{tabular}
\label{RV}
\end{table*}

\section{Peculiarities of the optical spectrum of HD\,56126}\label{results}

Optical spectra of PPNe differ from the spectra of classical supergiants by the
anomalous profiles of spectral lines (H{\sc i}, Na{\sc i}, He{\sc i}), and
primarily, by the anomalous H$\alpha$ profiles. H$\alpha$ lines in the spectra
of typical PPNe have complex emission and absorption profiles with asymmetric
cores, P\,Cyg- or inverse P\,Cyg-type profiles, and profiles with two emission
components. PPNe often exhibit a combination of several such features. Emission
in H$\alpha$ may be due to mass outflow and/or pulsations and hence we must
observe sporadic stellar wind in many PPNe. The Doppler shift of the core is
usually smaller than the escape velocity, i.e., we have evidence only for
motions at the wind base. The spectra of individual objects owe the great
variety of their profiles to the differences in the dynamical processes  in
their extended atmospheres: spherically symmetric outflow with constant or
height-dependent velocity, mass infall onto the photosphere, and pulsations. A
two-component emission profile is indicative of a nonspherical envelope, e.g.,
the presence of a circumstellar disk.

The peculiarity of the optical spectra of PPNe often shows up not only in
specific H{\sc i} profiles, but also in the distortions of the spectral
features of the $F-K$-type supergiant due to chemical composition anomalies and
the presence of molecular features along with atomic and ion lines.

\begin{figure}
\includegraphics[angle=0,width=0.9\textwidth,bb=20 20 320 200,clip=]{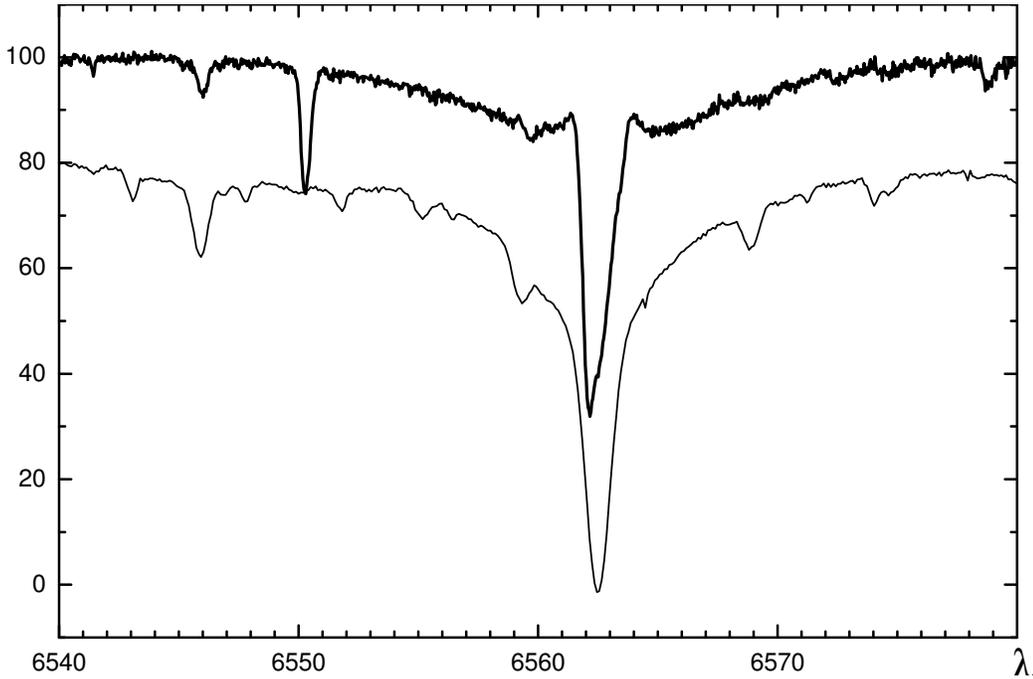}
\caption{Fragment of the atlas containing the H$\alpha$
         profile in the spectra of HD\,56126 (top) and $\alpha$\,Per
	 (bottom). The $y$-axis gives the residual intensities,
	 the continuum level is set equal to 100.}
\label{Halpha}
\end{figure}

\begin{figure}
\includegraphics[angle=0,width=0.4\textwidth,height=1.15\textwidth,bb=0 5 355 730,clip=]{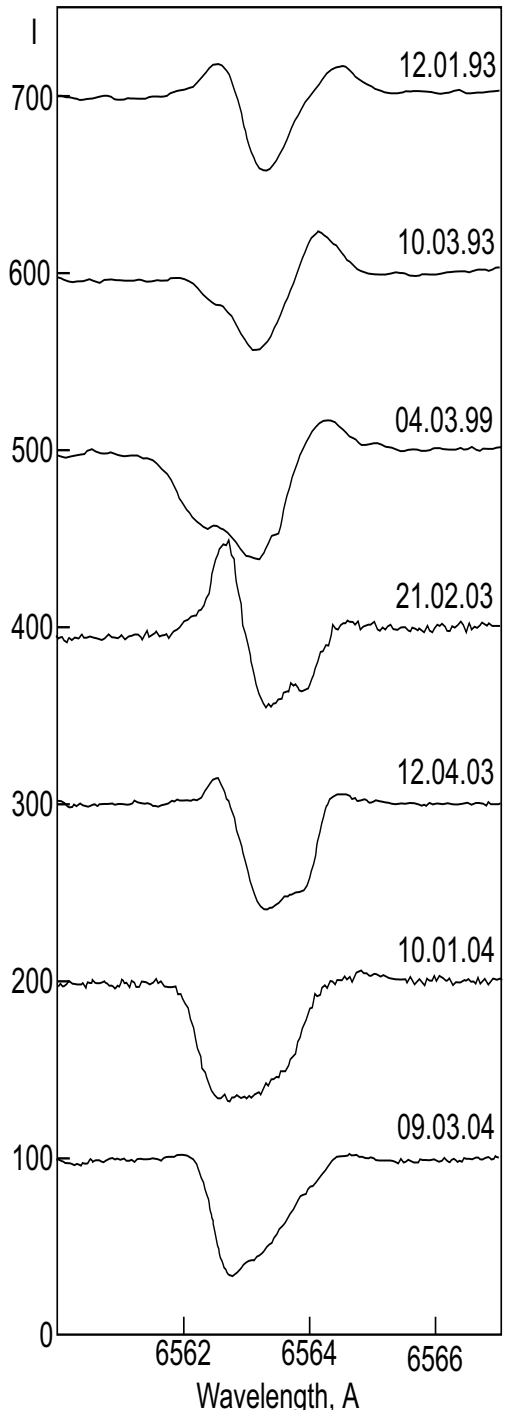}
\caption{Variations of the H$\alpha$-line profile in the
        spectra of HD\,56126 taken on different days.
	The $y$-axis shows the residual intensities, the continuum level
	of the lower spectrum is set to 100 and every next spectrum is
	shifted by 100 gradations with respect to the previous spectrum.}
\label{Halpha-var}
\end{figure}

HD\,56126 exhibits all these spectral peculiarities that distinguish PPN
from a normal supergiant of the same spectral type. As is evident from
Fig.\,\ref{Halpha}, the H$\alpha$ line has a complex profile with
absorption and emission components, which are absent in the spectrum of
the comparison star $\alpha$\,Per. Figure\,\ref{Halpha} also shows
well-defined photospheric wings of the H$\alpha$ line in the spectrum of
HD\,56126. These wings are almost as extended as in the spectrum of
$\alpha$\,Per. Figure\,\ref{Halpha-var}, which shows all the data now
available, demonstrates date-to-date variations of the central part of the
H$\alpha$ profile. Earlier, Oudmaijer and Bakker [\cite{OudBakk}] performed
spectral monitoring of HD\,56126 and also found the H$\alpha$ to be highly
variable over a two-months time scale. H$\alpha$-line variability can be
naturally explained in the case of post-AGB stars with signs of binarity
(e.g., in the case of HR\,4049 [\cite{Bakk98}]), however, it also shows up
in post-AGB objects, which exhibit no regular radial-velocity or light
variations (the case of HD\,133656 [\cite{Winck96}]). Photometric
variability would allow us (like in the case of RV\,Tau type stars) to
invoke the mechanism where a shock wave stimulates mass outflow. Based on
an extensive set of good quality spectroscopic observations of HD\,56126,
Barth\`es et al. [\cite{Barthes}] found that not only the profile of
H$\alpha$ but also that of H$\beta$ to be variable. The above authors
analyzed the variations of the profiles of both these lines and concluded
that no periodic component is present that could be associated with the
radial-velocity and photometric variations of the star.

\begin{figure}
\includegraphics[angle=0,width=0.9\textwidth,bb=10 10 310 200,clip=]{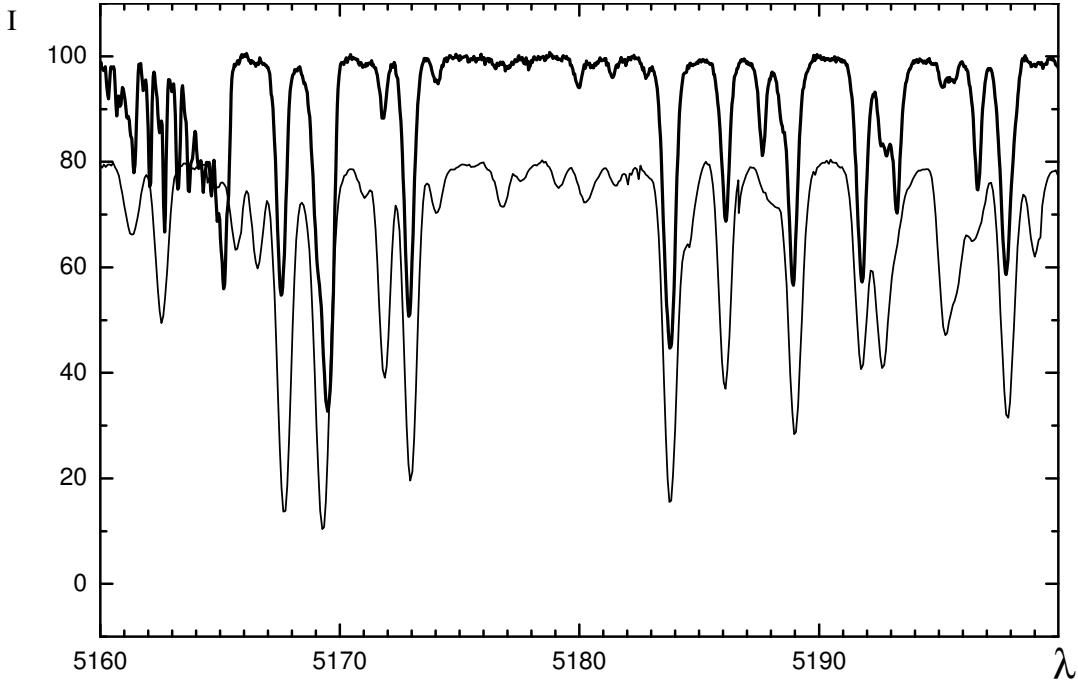}
\caption{Same as Fig.\,\ref{Halpha}, but for the spectral region
        containing the  5165\,\AA{} band of the Swan system of C$_2$
	molecule and the Fe{\sc ii}(42) 5169\,\AA{} line.}
\label{Swan}
\end{figure}

The profiles of strong Fe{\sc ii} lines (first and foremost, those of the
members of the 42-nd multiplet), Ba{\sc ii}, and other elements in the
spectrum of HD\,56126 are also variable. However, whereas either the blue
or red wing may have lower slopes in the absorption core of H$\alpha$,
nonhydrogen absorptions preserve their asymmetry pattern unchanged: the
blue wing is always more extended than the red wing. The profile of the
Fe{\sc ii}\,(42)\,$\lambda$\,5169\,\AA{} line in Fig.\ref{Swan} is a
typical example.

\begin{figure}
\includegraphics[angle=0,width=0.9\textwidth,bb=10 10 310 200,clip=]{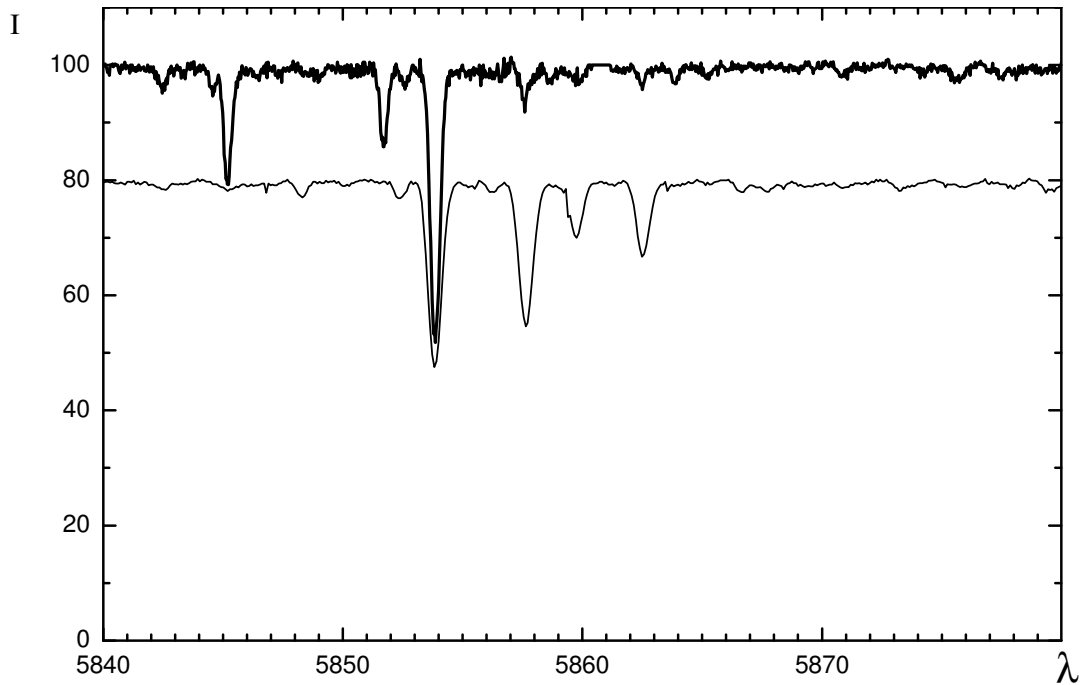}
\caption{Same as Fig.\,\ref{Halpha}, but for the spectral region
         with the  Ba{\sc ii} $\lambda$\,5853\,\AA{} line.}
\label{atlas52}
\end{figure}

Figures \,\ref{Swan} and \ref{atlas52} show fragments of the spectra that
may illustrate the differences between line intensities of HD\,56126 and
$\alpha$\,Per. Fe{\sc ii} absorptions in the spectrum of the former star
are much weaker than in those of the latter star, and the ratio of the
central depths of the same line in the spectra of the two stars depends on
the line intensity: it increases from 1.5 to 4 as one passes from weak to
strong lines. Fe{\sc i} lines are also depressed, on the average, by 0.1.
On the contrary, C{\sc i} absorptions as well as those of Y{\sc ii},
Zr{\sc ii}, and other $s$-process products are deeper by 0.1--0.2 in the
spectra of HD\,56126 compared to the corresponding features in the
spectrum of $\alpha$\,Per.

Let us now analyze the molecular component of the spectrum of  HD\,56126.
Bakker et al. [\cite{Bakk96}] were the first to identify the Swan
absorption bands of C$_2$ molecule and of the red system of CN molecule in the
spectrum of the star. Later, Bakker et al. [\cite{Bakk97}] used
high-resolution spectra with $R$=50000 to perform a detailed analysis of
molecular bands in the spectra of HD\,56126 and 16 other PPNe selected based on
the presence of carbon molecules C$_2$, CN, CH$^+$ in their shells. Judging by
the velocity corresponding to the position of these bands, the molecular
spectrum forms in a limited region of the shell close to the star
[\cite{Bakk97}]. Our spectra exhibit several bands of the Swan system
(see Fig.\,\ref{Swan}).

Here it is pertinent to recall that the spectra of several PPNe were found
to exhibit emission Swan bands [\cite{Bakk96,Kloch98}]. However, the spectra
of HD\,56126 taken in different years show no signs of emission in these
bands. D lines of Na{\sc i} neither show any signs of emission. This fact
is consistent with the rather simple elliptical shape of the nebula
surrounding HD\,56126. Emissions in the Swan bands or in Na{\sc i} D lines
appear to show up only in the spectra of PPNe with bright circumstellar
nebulae with well-defined asymmetry. The spectroscopy of the following PPNe
corroborates this hypothesis: IRAS\,04296$+$3429 [\cite{IRAS04296}],
IRAS\,23304$+$6147 [\cite{IRAS23304}], AFGL\,2688 [\cite{Egg}],
IRAS\,08005$-$2356 [\cite{IRAS08005}], IRAS\,20056$+$1834 [\cite{Rao}], and
IRAS\,20508$+$2011 [\cite{IRAS20508}]. On HST images [\cite{Ueta}], the
nebulae surrounding these PPNe are usually asymmetric and have a bipolar
structure. Note also that most of the objects listed above belong to type
``1'' according to the classification of Trammell et al. [\cite{Trammell}]
--- i.e., they are PPNe with polarized optical radiation.

Figure\,\ref{NaD} shows the behavior of the D2 Na{\sc i} line profile in
the spectrum of HD\,56126, and here, like in Table\,\ref{RV}, to reveal
the fine structure of lines, we analyze only the spectra taken with the
highest resolution. The positions of the three short-wavelength components
remain constant within the errors. This stability confirms that the
components form in the interstellar medium. The wavelength shift of the
deepest component agrees with that of the Swan bands (columns 8 and 9 in
Table\,\ref{RV}), and this fact indicates that the component in question
forms in the circumstellar shell. Finally, the longest-wavelength
component forms in the photosphere: its temporal behavior agrees with that
of other photospheric absorptions (columns 4 and 8 in Table\,\ref{RV}).

\section{Radial velocities pattern}\label{RV-var}

Much attention has been paid to the radial-velocity variations of
HD\,56126 and to study the differences between the radial velocities
inferred from lines of different types. Bujarrabal et al. [\cite{Bujar}]
used CO millimeter-wave observations to find $V_r=86.1$\,km/s, which is
natural to adopt as the systemic radial velocity of HD\,56126. Based on an
extensive collection of spectrograms with high temporal resolution and
high $S/N$ ratio, Oudmaijer and Bakker [\cite{OudBakk}\ analyzed the
behavior of $V_r$ and concluded that it is variable on a time scale of
several months with a small amplitude ($V_r$=84$\div 87 \pm 2$\,km/s). The
above authors demonstrated the absence of variations on time scales
ranging from several minutes to several hours. The variability of the
radial velocity of HD\,56126 also showed up when the radial-velocity
measurements made with the 6-m telescope were compared to published data
[\cite{Kloch95}]. L\`ebre et al. [\cite{LebrMaur}] performed a detailed
spectroscopic monitoring of HD\,56126. Fourier analysis of the available
radial-velocity and photometric data led the above researchers to conclude
that the dynamical state of the atmosphere of HD\,56126 is similar to that
of the atmospheres of RV\,Tau-type variables. The above authors
interpreted the variability of H$\alpha$ in terms of shock propagation.
Later, L\`ebre et al. [\cite{LebrFok}] analyzed the variability of two
lines, H$\alpha$ and H$\beta$. They obtained additional spectroscopic data
and determined the period of radial pulsations to be $P=36.8$ days.

Barth\`es et al. [\cite{Barthes}] analyzed all the available reliable
radial-velocity measurements made for HD\,56126 (89 measurements over eight
years) and concluded that radial velocity of this star varies with a
half-amplitude of 2.7\,km/s and the main period of $P$ = 36.8 $\pm 0.2^d$. The
period of photometric variability is the same and photometric amplitude is very
small, $0\lefteqn{.}^m02$. However, the above authors found the variability
pattern of the star to differ significantly from pulsations observed in
RV\,Tau-type stars. Judging by its temperature [\cite{Kloch95}],
HD\,56126 evolved further beyond the stage of RV\,Tau-type stars. The
photometric and radial-velocity variations of HD\,56126 may be due to
first-overtone radial pulsations driven by shocks that generate complex
asynchronous motions in the upper hydrogen layers of the star.

Table\,\ref{RV} presents the radial-velocity data we obtained for
HD\,56126. Given that a velocity gradient is very likely in the upper
layers of the star's atmosphere, we report here the $V_r$ values for
individual lines and groups of lines. As is evident from Table\,\ref{RV},
velocity variations inferred from weak absorptions (their residual
intensities approach 1) are within the variability limits established by
Barth\`es et al. [\cite{Barthes}]. The positions of the circumstellar Na{\sc i} D
lines and Swan bands of the C$_2$ molecule agree well with each other and match
the expansion velocity of the shell, $V_{exp} \approx$11\,km/s. Note that the
position of the wind component of the H$\alpha$ line is inconsistent with those
of the wind components of other lines, whereas the variations of the
H$\beta$ profile is synchronized with those of Fe{\sc ii}(42) lines.

\begin{figure}
\includegraphics[angle=0,width=0.9\textwidth,bb=0 0 610 460,clip=]{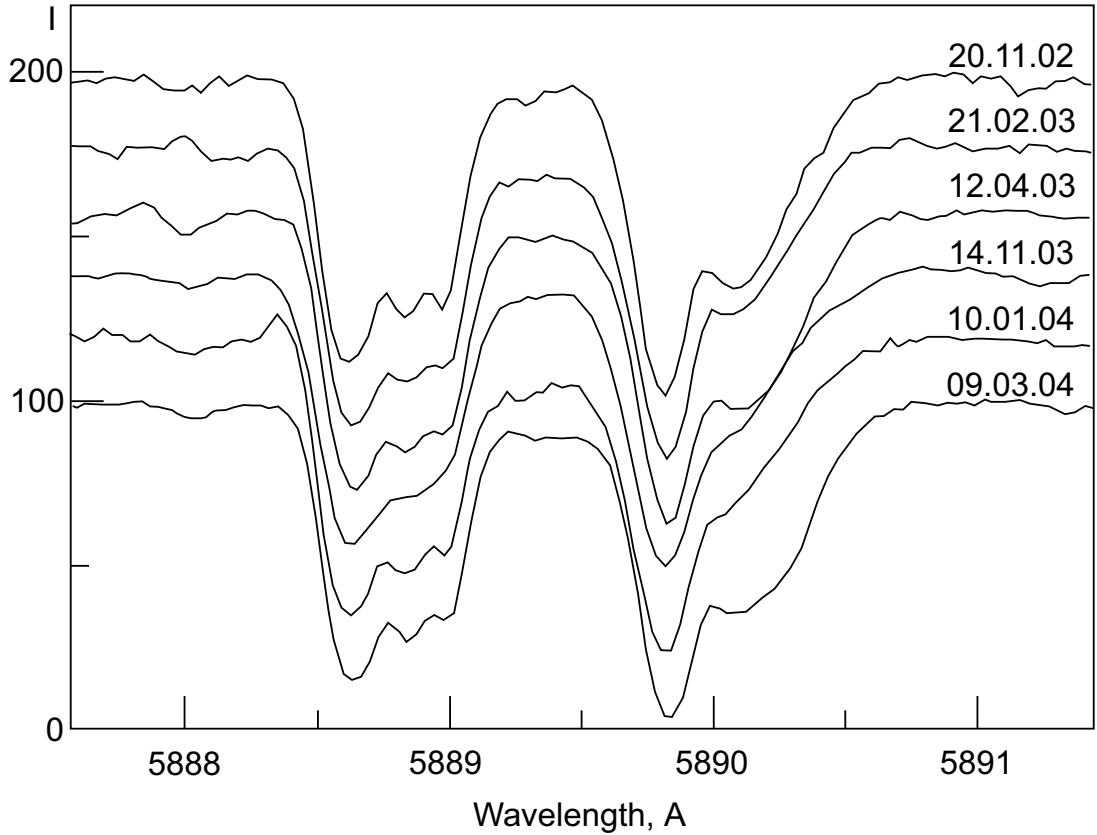}
\caption{Spectral fragment with the D2 Na{\sc i} line on different observing dates.}
\label{NaD}
\end{figure}

When intercomparing our data and comparing it with those of other authors one
must control the zero points of the corresponding radial-velocity systems. We
used interstellar and circumstellar lines to control the radial-velocity zero
points. Three blueshifted interstellar components of Na{\sc i} lines in the
spectrum of  HD\,56126, which are barely visible in
Fig.\,\ref{NaD}, yield $V_r$ values listed in Column 10--12 of
Table\,\ref{RV}. The fourth weak component with
$V_r\approx$46\,km/s is barely visible in at least three our spectra. For each
component, all our $V_r$ estimates agree with each other and with those of
Bakker et al. [\cite{Bakk96}] within the quoted errors. Furthermore,
as is evident from Fig.\,\ref{NaD}, the blend consisting of the
three main components has sharp boundaries, which allow the velocity of this
entire feature to be confidently measured. Its mean velocity as inferred from
our data is equal to $V_r=20.3\pm 0.3$\,km/s and agrees with the velocity of
20$\pm 2$\,km/s measured by L\`ebre et al. [\cite{LebrMaur}] from
lower-resolution spectra. Crawford and Barlow [\cite{Crawford}] showed
that when observed with superhigh resolution, the circumstellar  C$_2$ and
K{\sc i} features exhibit components that are about 1\,km/s apart. These
components yield the same set of velocity values, but have different
intensities. This effect may cause minor systematic differences (also on the
order of 1\,km/s) between the velocities inferred from circumstellar atomic and
molecular lines in lower-resolution spectra. Our measurements reveal no
variations of these velocities with time, and their mean values, 77.2$\pm 0.5$\
and 75.4$\pm 0.3$\,km/s for C$_2$ and Na{\sc i}, respectively, do not disagree
systematically with the radial velocities obtained by L\`ebre et al.
[\cite{LebrMaur}], Bakker et al. [\cite{Bakk96,Bakk97}], and Crawford and Barlow
[\cite{Crawford}]: 77.3$\div$77.6\ and 75.3$\div$76.8\,km/s for C$_2$
and Na{\sc i}, K{\sc i}, respectively.

However, when comparing our $V_r$ values with published data, one must take
into account not only methodological effects, but also the spectroscopic
peculiarity of the object itself. Line profiles in the spectrum of
HD\,56126 are asymmetric and their shape varies both with time and with
line intensity. We plan to undertake a detailed analysis of the velocity
field at different depths in the atmosphere and in the circumstellar
envelope of HD\,56126 in a separate paper.

As for the possible binarity of HD\,56126, it has neither been confirmed or
finally disproved. In this connection, of certain interest is a comment by
Barth\`es et al. [\cite{Barthes}] who pointed out a weak trend in the
star's radial velocity over several years of observations. This trend maybe
indicative of a second companion in the system with an orbital period longer
than 16 years. Our eight spectra, which we took on more recent dates, fail to
clarify the situation. It would be therefore important to follow the behavior
of $V_r$ over a several-years period taking one to two spectra every month on a
regular basis.

The radial-velocity variability of  HD\,56126 is not a unique phenomenon. Part
of candidate PPNe also demonstrate radial-velocity variations on a time scale
of several hundred days, which may be indicative of the binary nature of these
objects. Evidence for orbital motion has been obtained for several optically
bright objects with IR excess. For example, authors of papers
[\cite{Ferro,Waters93}] and [\cite{WaelLam}]
proved the binary nature, determined the orbital elements, and proposed models
for the high-latitude supergiants 89\,Her and HR\,4049, respectively. Van
Winckel et al. [\cite{Winck95}] showed that HR\,4049, HD\,44179, and
HD\,52961 to be spectral binaries with the orbital periods of about one to two
years. The above authors concluded that all extremely metal-deficient PPNe
studied so far (HR\,4049, HD\,44179, HD\,52961, HD\,46703, and BD\,+39$^{\rm
o}$4926) are binary stars. The observed correlation between the binarity and
the presence of a hot dust envelope indicates that binarity promotes the
formation of an envelope. Bakker et al. [\cite{Bakk98}] use
high-resolution spectra of HR\,4049 to analyze the variations of complex
emission--absorption profiles of  Na\ D lines and H$\alpha$ lines over the
orbital period. Individual components of these lines may form under different
conditions: in the atmosphere of the main star; in the disk where both
components of the binary are immersed, or in the interstellar medium. For such
binaries, of fundamental importance is the determination of the systemic
velocity from radio spectroscopy.

The nature of the companions of suspected binary post-AGB stars remains
unclear, because we see no direct manifestations of these companions
either in the continuum or in spectral lines (all known binary post-AGB
objects belong to type $SB1$). These companions may be either very hot
object, or main-sequence stars of very low luminosity. For example, Bakker
et al. [\cite{Bakk98}] believe that the secondary companion in HR\,4049 is
a cold (${\rm T_e=3500}$\,K) MS star with a mass of
$\mathcal{M}$=0.56$\mathcal{M}_{\odot}$, although it may also be a white
dwarf, like in the case of Ba-stars [\cite{McClure}].

Unfortunately, because of the short history of PPN studies we are so far unable
to make any definitive conclusions concerning the cause of radial-velocity
variability of a representative sample of these objects. Moreover, the observed
pattern of radial-velocity variations is often complicated by differential
motions in the extended atmospheres of these objects. A detailed analysis of
radial velocities based on data taken with high spectral and temporal
resolution for selected --- the brightest --- PPNe reveals differences in the
behavior of radial velocities inferred from lines of different excitation,
which form at different depths in the atmosphere of the star. For example,
Bakker et al. [\cite{BakkLam}] analyzed the spectrum of the IRAS
source identified with the peculiar supergiant HD\,101584 and found eight
categories of spectral lines with fundamentally different temporal behavior of
profiles, half-widths, and shifts (and hence of $V_r$ values). In particular,
the highest-excitation absorption features, which form near the star's
photosphere, exhibit radial-velocity variations due to the orbital motion in
the binary system. At the same time, low-excitation lines with P\,Cyg-type
profiles form in the stellar-wind region and are indicative of mass outflow.
The systemic velocity has been confidently determined from radio emissions of
CO and OH molecules.

\begin{figure}
\includegraphics[angle=0,width=0.9\textwidth,bb=10 10 310 200,clip=]{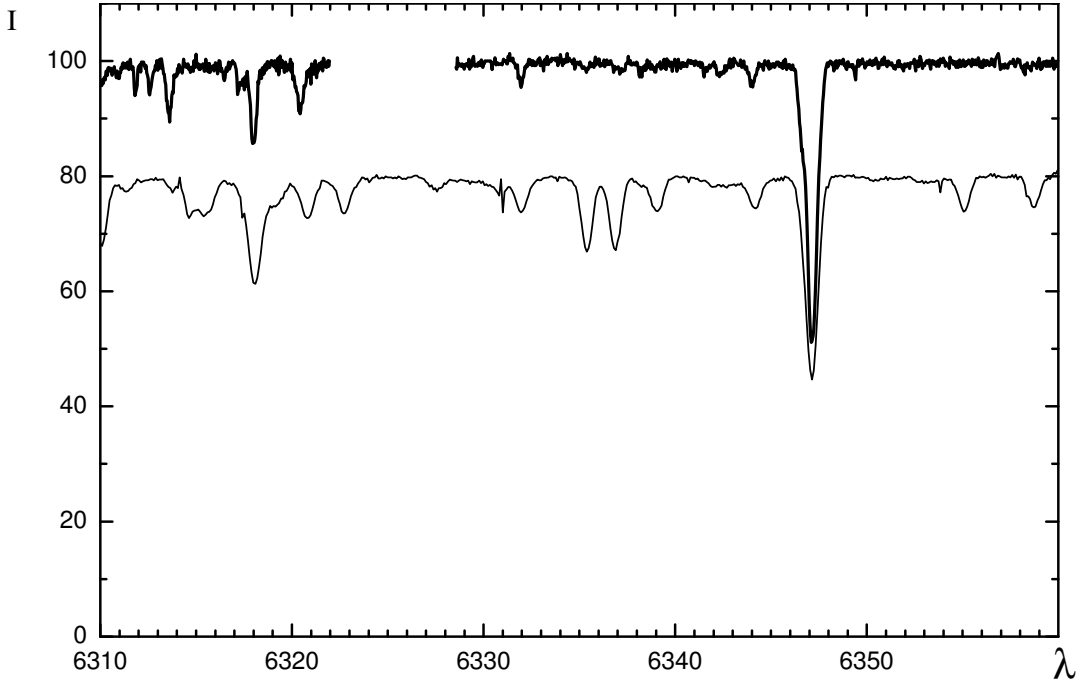}
\caption{Same as Fig.\,\ref{Halpha}, but for spectral fragment containing the Si{\sc ii}
              $\lambda$\,6347\,\AA\ line.}
\label{atlas65}
\end{figure}

\section{Spectral atlas}\label{atlas}

\begin{table}[hbpt]
\caption{Spectra used to create this atlas}
\medskip
\begin{tabular}{c|c|c|c|c}
\hline
& \multicolumn{2}{c}{\small  $\alpha$\,Per } & \multicolumn{2}{|c}{\small  HD\,56126} \\ \cline{2-5}
$\Delta\lambda$,  & Date & Spect- & Date & Spect-\\  
 \AA{} &&rograph  &      & rograph \\ 
\hline 4010--5460      & 11.11.05   & NES  &  12.11.05   & NES  \\  
5460--6010      & 12.11.05   & NES  &  9.03.04    & NES  \\  
6010--6640      &  4.03.99   & Lynx &  9.03.04    & NES  \\  
6640--8790      &            &      &  10.03.93   & Lynx \\ 
\hline
\end{tabular}
\label{fragments}
\end{table}

Our comparative atlas of the spectra of  HD\,56126 and $\alpha$\,Per includes
94 plots representing 40\,\AA{}-long spectral fragments. Some of them are shown
in Figs.\,\ref{Halpha}, \ref{Swan},
\ref{atlas52} and \ref{atlas65} to illustrate the
differences between the intensities and profiles of lines in the spectra of two
stars of similar temperature and luminosity. The full version of
 the  atlas is available at $http://www.sao.ru/hq/ssl/Atlas/Atlas.html$.

In the wavelength interval 4010--6640\,\AA\AA{} the atlas gives the complete
spectra of both objects. However, in the more long-wavelength portion, up to
8790\,\AA{}, part of the spectrum has been lost in gaps between echelle orders
and the remaining portions are overcrowded by telluric lines. The atlas
therefore gives only the most informative fragments for this part of the
spectrum of HD\,56126.

The spectrum of HD\,56126 is variable  --- line profiles, differential line
shifts, and radial velocities vary from date to date, and therefore we
performed no averaging of any kind --- different spectral intervals are
represented in the atlas by different spectra indicated in
Table\,\ref{fragments}. For each wavelength interval,  we selected
from among the available spectra the one with the highest resolution and
signal-to-noise ratio.

We supplement graphic data in the atlas with tables. In Table\,\ref{lines}
column~1 gives the results of identification of spectral features;
column~2, the laboratory wavelengths used to measure the radial
velocities; columns 3 and 5, the central residual line intensities ``r'',
and columns~4 and 6, the heliocentric radial velocities $V_r$ measured
from the line cores.

To identify atomic and molecular lines in the spectrum of HD\,56126, we
use the atlases and tables of solar spectrum [\cite{sun-atlas,Wallace,
Pierce,Swens}], the Moore tables for multiplets [\cite{Moore,Moore2}], and
electronic tables to the paper by Bakker et al. [\cite{Bakk96}].
We also use VALD database [\cite{VALD}]. We supplement the
standard identification criteria (wavelength, relative line intensity,
specific line profile) by two additional ones. One of these new criteria
uses the chemical composition anomalies of HD\,56126 mentioned above and
the spectrum of the comparison star. The second new criterion can be
applied only to sufficiently strong lines ($r<$\,0.5), which in some of
our spectra exhibit a sharp variation of radial velocity with line depth.
Several rather strong lines remained unidentified in the spectrum of
HD\,56126. Some of them can be seen in the fragments of the atlas
presented here: the $\lambda$\,6550\,\AA{} line in
Fig.\ref{Halpha}, the $\lambda$\,5845 and 5852\,\AA{} lines in
Fig.\,\ref{atlas52}, and the $\lambda$\,6347\,\AA{} line in
Fig.\,\ref{atlas65}.

Compared to the lines in the spectrum of  $\alpha$\,Per, those in the spectrum
of HD\,56126 are less blended, because they are narrower and, in addition, many
of these lines are weaker due to low metallicity of the star. However, by no
means all absorptions can be used for reliable measurement of radial
velocities. Table\,\ref{lines} lists about one and half thousand
identifications for both stars and only  940  $V_r$ measurements for HD\,56126
obtained mostly from lines with minimal blending or from lines with the
strongest difference of intensity in the spectra of two stars.

\section*{Conclusions}

We use numerous high-resolution observations ($R$=25000 and 60000) made with the
echelle spectrographs of the 6-m telescope to perform a detailed analysis of
the optical spectrum of the  post-AGB star HD\,56126 identified with the IR
source IRAS\,07134$+$1005. We identified  numerous absorptions of neutral atoms
and ions in the wavelength interval from 4010 to 8790\,\AA\AA{} and measured
their depths and the corresponding radial velocities. We identified absorption
bands of the C$_2$, CN, and CH molecules, and interstellar bands (DIB). In
addition to the known variability of the profile of the  H$\alpha$ line, we
found variations of the profiles of a number of Fe{\sc ii} and Ba{\sc ii}
lines. We produced an atlas of the spectra of HD\,56126 and its comparison star
$\alpha$\,Per.

An analysis of our radial velocities determined from all spectra of our
collection leads us to conclude that:
\begin{itemize}
\item{} the accuracy of our radial-velocity data for HD\,56126 allows them to
be combined with the most accurate of earlier published measurements;

\item{} we found the behavior of  $V_r$ values to differ for lines of different
excitation degree, which form at different depth in the stellar atmosphere. The
half-amplitude of the variations of radial velocities measured from weak
absorptions  ({\it r}\,$\rightarrow$\,1) is equal to  2--3\,km/s;

\item{} we confirm the stability of the expansion velocity of the circumstellar
envelope of HD\,56126 as measured from  C$_2$ and Na{\sc i} lines;

\item{} we reveal the complex and variable shape of the profiles of strong
lines (not only hydrogen lines, but also absorption features of Fe{\sc ii},
Y{\sc ii}, Ba{\sc ii}, and other elements), which form in the expanding
atmosphere (wind base) of the star. To study the kinematic state of the
atmosphere, one needs measurements of radial velocities for individual
details of these profiles;

\item{} we demonstrate the necessity of high and even superhigh spectral
resolution for studying stellar and circumstellar lines, respectively, in the
spectrum of  HD\,56126.
\end{itemize}

\section*{Acknowledgments}
This work is supported by the Russian Foundation for Basic Research (project
code 05--07--90087), the Presidium of the Russian Academy of Sciences (program
``Origin and Evolution of Stars and Galaxies``, the Branch of Physical Sciences
of the Russian Academy of Sciences (program ``Extended Objects in the
Universe''). This publication is based on work supported by Award No.
RUP1--2687--NA--05 of the U.S.
 Civilian Research and Development Foundation (CRDF).

This work makes use of the SIMBAD, NASA ADS, and VALD astronomical
databases.

{}

\clearpage
\newpage

\tablecaption{List of lines identified in the spectra of  HD\,56126 and
         $\alpha$\,Per. Columns 3 and 5 list the central residual
	 intensities of the lines (the continuum level is set to  1),
	 and columns 4 and 6, the heliocentric velocities $V_r$.}
\tablehead{\hline & & \multicolumn{2}{c}{\underline{\qquad$\alpha$\,Per \qquad}} & \multicolumn{2}{c}{\underline{\quad HD\,56126}} \\    
Element &  $\lambda$\,\AA{}&\quad r & $V_r$ &\quad  r & $V_r$  \\
 [2pt] \hline }
\tabletail{\hline \rule{0pt}{5pt}&&&&& \\} 
\begin{supertabular}{l l @{\quad} l c@{\quad}  l @{\quad}l @{\quad}l@{\quad} }
TiII(11)      & 4012.39 & 0.13: & $-$2.2  & 0.05  & 81.7 \\
FeI(557)      & 4013.64 &       &         &       &      \\
FeI(485)      & 4013.82 & 0.53: &         &       &      \\
ScII(8)       & 4014.53 & 0.32: & $-$3.5  & 0.49  &      \\
CeII(157)     & 4014.90 &       &         & 0.56  &      \\
NiII(12)      & 4015.47 & 0.50: &         & 0.64  & 84.8:\\
CeII(256)     & 4015.88 &       &         & 0.74: & 83.0:\\
FeI(560)      & 4016.42 & 0.74: & $-$1.0  &       &      \\
\hline
\end{supertabular}
\label{lines}


\newpage

\begin{thebibliography}{}

\bibitem{Block} 1. T.~Bl\"ocker, Astrophys. J. {\bf 299}, 755 (1995).

\bibitem{Kwok} 2. S.~Kwok, Annu. Rev. Astron. \& Astrophys.  {\bf 31}, 63 (1993).

\bibitem{Ueta} 3. T.~Ueta, M.~Meixner, and M.~Bobrowsky, Astrophys. J. {\bf 528}, 861 (2000).

\bibitem{Crawford} 4. I.A.~Crawford and M.J.~Barlow, MNRAS {\bf 311}, 370 (2000).

\bibitem{Kloch95} 5. V.G.~Klochkova, MNRAS \textbf{272}, 710 (1995).

\bibitem{Kloch98} 6. V.G.~Klochkova, Bull. Spec. Astrophys. Observ. \textbf{44}, 5 (1998).

\bibitem{Decin} 7. L.~Decin, H.~van Winckel, C.~Waelkens, and E.~Bakker,  Astron. \& Astrophys. 
        \textbf{332}, 928 (1998).

\bibitem{nes1} 8. V.E.~Panchuk, V.G.~Klochkova, I.D.~Najdenov, Preprint of
the Special Astrophysical Observatory of the Russian Academy of Sciences
            No.\,135 (1999).

\bibitem{nes2} 9. V.E.~Panchuk, N.E.~PIskunov, V.G.~Klochkova, et al.,
Preprint of the Special Astrophysical Observatory of the Russian Academy
of Sciences No.\,169 (2002).

\bibitem{lynx} 10. V.G.~Klochkova, S.V.~Ermakov, V.E.~Panchuk, et al.,
Preprint of the Special Astrophysical Observatory of the Russian Academy
of Sciences No.\,137 (1999).

\bibitem{lynx2} 11. V.E.~Panchuk, V.G.~Klochkova, I.D.~Najdenov, et al.,
         Preprint of the Special Astrophysical Observatory of the Russian Academy
         of Sciences No.\,139 (1999).

\bibitem{slicer} 12. V.E.~Panchuk, M.V.~Yushkin, I.D.~Najdenov, Preprint of the Special
           Astrophysical Observatory of the Russian Academy of Sciences No.\,179
          (2003).

\bibitem{ECHELLE} 13. M.V.~Yushkin, V.G.~Klochkova, Preprint of the Special Astrophysical
          Observatory of the Russian Academy of Sciences, No.\,206 (2005).

\bibitem{gala} 14. G.A.~Galazutdinov, Preprint of the Special Astrophysical Observatory of
the Russian Academy of Sciences No.\,192 (1992).

\bibitem{OudBakk} 15. R.~Oudmaijer and E.J.~Bakker, MNRAS \textbf{271}, 615
(1994).

\bibitem{Bakk98} 16. E.J.~Bakker, D.L.~Lambert, H.~van Winckel, et al.,
        Astron. \& Astrophys. \textbf{336}, 263 (1998).

\bibitem{Winck96} 17. H.~van Winckel, R.~Oudmaijer, and N.R.~Trams., Astron. \& Astrophys. 
\textbf{312}, 553 (1996).

\bibitem{Barthes} 18. D.~Barth\`es, A.~L\`ebre, D.~Gillet, and N.~Mauron, Astron. \& Astrophys. 
\textbf{359}, 168 (2000).

\bibitem{Bakk96} 19. E.J.~Bakker, L.B.F.M.~Waters, H.J.G.L.M.~Lamers,
et.al., Astron. \& Astrophys.  \textbf{310}, 893 (1996).

\bibitem{Bakk97} 20. E.J.~Bakker, E.F.~Dishoeck, L.B.F.M.~Waters, and T.~Schoenmaker,
       Astron. \& Astrophys.  \textbf{323}, 469 (1997).

\bibitem{IRAS04296} 21. V.G.~Klochkova, R.~Szczerba, V.E.~Panchuk, and
              K.~Volk, Astron. \& Astrophys.  \textbf{345}, 905 (1998).  

\bibitem{IRAS23304} 22. V.G.~Klochkova, R.~Szczerba, V.E.~Panchuk, Astron.
            Lett. {\bf 26}, 88 (2000).

\bibitem{Egg} 23. V.G.~Klochkova, R.~Szczerba, V.E.~Panchuk, Astron. Lett. {\bf 26}, 439
                      (2000)

\bibitem{IRAS08005} 24. V.G.~Klochkova, E.L.~Chentsov, Astron. Rep. {\bf 48}, 301 (2004)

\bibitem{Rao} 25. N.~Kameswara Rao, A.~Goswami, and D.L.~Lambert, MNRAS
              \textbf{334}, 129 (2002).

\bibitem{IRAS20508} 26. V.G.~Klochkova, V.E.~Panchuk, N.S.~Tavolganskaya, G.~Zhao,
                   Astron. Rep. \textbf{83}, 265 (2006).

\bibitem{Trammell} 27. S.~Trammell, H.L.~Dinerstein, and R.W.~Goodrich, Astrophys. J.
                  \textbf{108}, 984  (1994).

\bibitem{Bujar} 28. V.~Bujarrabal, J.~Alcolea, and P.~Planesas, Astron. \& Astrophys.  \textbf{257}, 701 (1992).

\bibitem{LebrMaur} 29. A.~L\`ebre, N.~Mauron, D.~Gillet, and D.~Barthes, Astron. \& Astrophys.  \textbf{310},
                  923 (1996).

\bibitem{LebrFok} 30. A.~L\`ebre, A.~B.~Fokin, D.~Barthes, D.~Gillet, and N.~Mauron,
             Astrophys. Space Sci.\ \textbf{265}, 105 (2001).

\bibitem{Ferro} 31. A.A.~Ferro, PASP \textbf{96}, 641 (1984).

\bibitem{Waters93} 32. L.B.F.M.~Waters, C.~Waelkens, M.~Mayor, and N.R.~Trams,
         Astron. \& Astrophys.  \textbf{269}, 242 (1993).

\bibitem{WaelLam} 33. C.~Waelkens, H.J.G.L.M.~Lamers, L.B.F.M.~Waters, et al.,
                  Astron. \& Astrophys.  \textbf{242}, 433 (1991).

\bibitem{Winck95} 34. H.~van Winckel, C.~Waelkens, and L.B.F.M.~Waters,
               Astron. \& Astrophys.  \textbf{293}, L25
                  (1995).

\bibitem{McClure} 35. R.D.~McClure,  MNRAS \textbf{96}, 117 (1984).

\bibitem{BakkLam} 36. E.~J.~Bakker,  H.J.G.L.M.~Lamers,  L.B.F.M.~Waters,
                et al., Astron. \& Astrophys.  \textbf{307}, 869  (1996).

\bibitem{sun-atlas} 37. R.L.~Kurucz, I.~Furenlid, and J.T.L.~Brault, Nat. Solar Observ. Atlas,
                    New Mexico: National Solar Observatory (1984).

\bibitem{Wallace} 38. L.~Wallace, K.~Hinkle, and W.~Livingston, Nat. Solar Obs. Techn. Rep.
 No.00--001, Tucson (2000).

\bibitem{Pierce} 39. A.K.~Pierce and J.B.~Breckinridge, Contr. Kitt Peak Nat. Obs.\ No.\,559 (1973).

\bibitem{Swens} 40. J.W.~Swensson, W.S.~Benedict, L.~Delbouille, and G.~Roland,
        ``The solar spectrum from $\lambda$~7498 to $\lambda$~12016.
	A table of measures and identifications'',
  Mem. Soc. Roy. Sci. Liege, Vol.hors ser. No.5 (1970).

\bibitem{Moore} 41. C.E.~Moore, Contrib. Princeton University Observ.\ No.{20}, part I (1945).

\bibitem{Moore2} 42. C.E.~Moore, Contrib. Princeton University Observ. No.{20}, part II (1945).

\bibitem{VALD} 43. N.E.~Piskunov, F.~Kupka, and T.A.~Ryabchikova, A\&AS \textbf{112}, 525 (1995).

\end{thebibliography}
\end{document}